\documentclass[aps,prl,twocolumn,superscriptaddress,showpacs,amsmath,amssymb]{revtex4}

\usepackage{graphicx} 
\usepackage{dcolumn}  
\usepackage{rotating}
\usepackage{ulem}
\usepackage{threeparttable}
\usepackage[colorlinks,linkcolor=red,anchorcolor=green,citecolor=blue]{hyperref}

\newcommand{\BR}{{\cal B}}
\newcommand{\EE}{e^+e^-}

\newcommand {\tabincell}[2]{\begin{tabular}{@{}#1@{}}#2\end{tabular}}%

\begin{document}
\hyphenpenalty=10000
\vspace*{-3\baselineskip}


\preprint{\vbox{ \hbox{   }
                       \hbox{Intended for {\it P.R.L}}
                       \hbox{Author: Y. B. Li, C. P. Shen}
                       \hbox{Committee: J. Yelton(chair),}
                       \hbox{ $~~~~~~~~~~~~~~~~$B. Grube, U. Tamponi }
}
}

\title{ 
	First measurements of absolute branching fractions of the $\Xi_c^+$ baryon at Belle}


\noaffiliation
\affiliation{University of the Basque Country UPV/EHU, 48080 Bilbao}
\affiliation{Beihang University, Beijing 100191}
\affiliation{Brookhaven National Laboratory, Upton, New York 11973}
\affiliation{Budker Institute of Nuclear Physics SB RAS, Novosibirsk 630090}
\affiliation{Faculty of Mathematics and Physics, Charles University, 121 16 Prague}
\affiliation{Chonnam National University, Kwangju 660-701}
\affiliation{University of Cincinnati, Cincinnati, Ohio 45221}
\affiliation{Deutsches Elektronen--Synchrotron, 22607 Hamburg}
\affiliation{University of Florida, Gainesville, Florida 32611}
\affiliation{Key Laboratory of Nuclear Physics and Ion-beam Application (MOE) and Institute of Modern Physics, Fudan University, Shanghai 200443}
\affiliation{Justus-Liebig-Universit\"at Gie\ss{}en, 35392 Gie\ss{}en}
\affiliation{Gifu University, Gifu 501-1193}
\affiliation{II. Physikalisches Institut, Georg-August-Universit\"at G\"ottingen, 37073 G\"ottingen}
\affiliation{SOKENDAI (The Graduate University for Advanced Studies), Hayama 240-0193}
\affiliation{Hanyang University, Seoul 133-791}
\affiliation{University of Hawaii, Honolulu, Hawaii 96822}
\affiliation{High Energy Accelerator Research Organization (KEK), Tsukuba 305-0801}
\affiliation{J-PARC Branch, KEK Theory Center, High Energy Accelerator Research Organization (KEK), Tsukuba 305-0801}
\affiliation{Forschungszentrum J\"{u}lich, 52425 J\"{u}lich}
\affiliation{IKERBASQUE, Basque Foundation for Science, 48013 Bilbao}
\affiliation{Indian Institute of Science Education and Research Mohali, SAS Nagar, 140306}
\affiliation{Indian Institute of Technology Guwahati, Assam 781039}
\affiliation{Indian Institute of Technology Hyderabad, Telangana 502285}
\affiliation{Indian Institute of Technology Madras, Chennai 600036}
\affiliation{Indiana University, Bloomington, Indiana 47408}
\affiliation{Institute of High Energy Physics, Chinese Academy of Sciences, Beijing 100049}
\affiliation{Institute of High Energy Physics, Vienna 1050}
\affiliation{Institute for High Energy Physics, Protvino 142281}
\affiliation{INFN - Sezione di Napoli, 80126 Napoli}
\affiliation{INFN - Sezione di Torino, 10125 Torino}
\affiliation{Advanced Science Research Center, Japan Atomic Energy Agency, Naka 319-1195}
\affiliation{J. Stefan Institute, 1000 Ljubljana}
\affiliation{Institut f\"ur Experimentelle Teilchenphysik, Karlsruher Institut f\"ur Technologie, 76131 Karlsruhe}
\affiliation{Kennesaw State University, Kennesaw, Georgia 30144}
\affiliation{Department of Physics, Faculty of Science, King Abdulaziz University, Jeddah 21589}
\affiliation{Kitasato University, Sagamihara 252-0373}
\affiliation{Korea Institute of Science and Technology Information, Daejeon 305-806}
\affiliation{Korea University, Seoul 136-713}
\affiliation{Kyoto University, Kyoto 606-8502}
\affiliation{Kyungpook National University, Daegu 702-701}
\affiliation{LAL, Univ. Paris-Sud, CNRS/IN2P3, Universit\'{e} Paris-Saclay, Orsay}
\affiliation{\'Ecole Polytechnique F\'ed\'erale de Lausanne (EPFL), Lausanne 1015}
\affiliation{P.N. Lebedev Physical Institute of the Russian Academy of Sciences, Moscow 119991}
\affiliation{Liaoning Normal University, Dalian 116029}
\affiliation{Faculty of Mathematics and Physics, University of Ljubljana, 1000 Ljubljana}
\affiliation{Ludwig Maximilians University, 80539 Munich}
\affiliation{Luther College, Decorah, Iowa 52101}
\affiliation{Malaviya National Institute of Technology Jaipur, Jaipur 302017}
\affiliation{University of Malaya, 50603 Kuala Lumpur}
\affiliation{University of Maribor, 2000 Maribor}
\affiliation{Max-Planck-Institut f\"ur Physik, 80805 M\"unchen}
\affiliation{School of Physics, University of Melbourne, Victoria 3010}
\affiliation{University of Mississippi, University, Mississippi 38677}
\affiliation{University of Miyazaki, Miyazaki 889-2192}
\affiliation{Moscow Physical Engineering Institute, Moscow 115409}
\affiliation{Moscow Institute of Physics and Technology, Moscow Region 141700}
\affiliation{Graduate School of Science, Nagoya University, Nagoya 464-8602}
\affiliation{Kobayashi-Maskawa Institute, Nagoya University, Nagoya 464-8602}
\affiliation{Universit\`{a} di Napoli Federico II, 80055 Napoli}
\affiliation{Nara Women's University, Nara 630-8506}
\affiliation{National Central University, Chung-li 32054}
\affiliation{National United University, Miao Li 36003}
\affiliation{Department of Physics, National Taiwan University, Taipei 10617}
\affiliation{H. Niewodniczanski Institute of Nuclear Physics, Krakow 31-342}
\affiliation{Nippon Dental University, Niigata 951-8580}
\affiliation{Niigata University, Niigata 950-2181}
\affiliation{Novosibirsk State University, Novosibirsk 630090}
\affiliation{Osaka City University, Osaka 558-8585}
\affiliation{Pacific Northwest National Laboratory, Richland, Washington 99352}
\affiliation{Panjab University, Chandigarh 160014}
\affiliation{State Key Laboratory of Nuclear Physics and Technology, Peking University, Beijing 100871}
\affiliation{University of Pittsburgh, Pittsburgh, Pennsylvania 15260}
\affiliation{Theoretical Research Division, Nishina Center, RIKEN, Saitama 351-0198}
\affiliation{University of Science and Technology of China, Hefei 230026}
\affiliation{Seoul National University, Seoul 151-742}
\affiliation{Showa Pharmaceutical University, Tokyo 194-8543}
\affiliation{Soongsil University, Seoul 156-743}
\affiliation{University of South Carolina, Columbia, South Carolina 29208}
\affiliation{Stefan Meyer Institute for Subatomic Physics, Vienna 1090}
\affiliation{Sungkyunkwan University, Suwon 440-746}
\affiliation{School of Physics, University of Sydney, New South Wales 2006}
\affiliation{Department of Physics, Faculty of Science, University of Tabuk, Tabuk 71451}
\affiliation{Tata Institute of Fundamental Research, Mumbai 400005}
\affiliation{Department of Physics, Technische Universit\"at M\"unchen, 85748 Garching}
\affiliation{Toho University, Funabashi 274-8510}
\affiliation{Department of Physics, Tohoku University, Sendai 980-8578}
\affiliation{Earthquake Research Institute, University of Tokyo, Tokyo 113-0032}
\affiliation{Department of Physics, University of Tokyo, Tokyo 113-0033}
\affiliation{Tokyo Institute of Technology, Tokyo 152-8550}
\affiliation{Tokyo Metropolitan University, Tokyo 192-0397}
\affiliation{Virginia Polytechnic Institute and State University, Blacksburg, Virginia 24061}
\affiliation{Wayne State University, Detroit, Michigan 48202}
\affiliation{Yamagata University, Yamagata 990-8560}
\affiliation{Yonsei University, Seoul 120-749}
  \author{Y.~B.~Li}\affiliation{State Key Laboratory of Nuclear Physics and Technology, Peking University, Beijing 100871} 
  \author{C.~P.~Shen}\affiliation{Key Laboratory of Nuclear Physics and Ion-beam Application (MOE) and Institute of Modern Physics, Fudan University, Shanghai 200443} 
  \author{I.~Adachi}\affiliation{High Energy Accelerator Research Organization (KEK), Tsukuba 305-0801}\affiliation{SOKENDAI (The Graduate University for Advanced Studies), Hayama 240-0193} 
  \author{J.~K.~Ahn}\affiliation{Korea University, Seoul 136-713} 
  \author{H.~Aihara}\affiliation{Department of Physics, University of Tokyo, Tokyo 113-0033} 
  \author{S.~Al~Said}\affiliation{Department of Physics, Faculty of Science, University of Tabuk, Tabuk 71451}\affiliation{Department of Physics, Faculty of Science, King Abdulaziz University, Jeddah 21589} 
  \author{D.~M.~Asner}\affiliation{Brookhaven National Laboratory, Upton, New York 11973} 
  \author{H.~Atmacan}\affiliation{University of South Carolina, Columbia, South Carolina 29208} 
  \author{T.~Aushev}\affiliation{Moscow Institute of Physics and Technology, Moscow Region 141700} 
  \author{R.~Ayad}\affiliation{Department of Physics, Faculty of Science, University of Tabuk, Tabuk 71451} 
% ...
\collaboration{The Belle Collaboration}

\begin{abstract}

We present the first measurements of the absolute branching fractions of $\Xi_c^+$ decays into $\Xi^- \pi^+ \pi^+$ and $p K^- \pi^+$ final states.  Our analysis is based on a data set of $(772\pm 11)\times 10^{6}$ $B\bar{B}$ pairs collected at the $\Upsilon(4S)$ resonance with the Belle detector at the KEKB $e^+e^-$ collider.
We measure the absolute branching fraction of $\bar{B}^{0} \to \bar{\Lambda}_{c}^{-} \Xi_{c}^{+}$
with the $\Xi_c^+$ recoiling against $\bar{\Lambda}_c^-$ in $\bar{B}^0$ decays resulting in $\BR(\bar{B}^{0} \to \bar{\Lambda}_{c}^{-} \Xi_{c}^{+}) = [1.16 \pm 0.42(\rm stat.) \pm 0.15(\rm syst.)] \times 10^{-3}$. We then measure the product branching fractions $\BR(\bar{B}^{0} \to \bar{\Lambda}_c^- \Xi_c^+)\BR(\Xi_c^+ \to \Xi^- \pi^+ \pi^+)$ and $\BR(\bar{B}^{0} \to \bar{\Lambda}_c^- \Xi_c^+)\BR(\Xi_c^+ \to p K^- \pi^+)$.
Dividing these product branching fractions by $\bar{B}^{0} \to \bar{\Lambda}_{c}^{-} \Xi_{c}^{+}$
yields:
$\BR(\Xi_c^+ \to \Xi^- \pi^+ \pi^+) = [2.86 \pm 1.21(\rm stat.) \pm 0.38(\rm syst.)]\%$ and
$\BR(\Xi_c^+ \to p K^- \pi^+)=[0.45 \pm 0.21(\rm stat.) \pm 0.07(\rm syst.)]\%$.
Our result for $\BR(\Xi_c^+ \to \Xi^- \pi^+ \pi^+)$ can be combined with $\Xi_c^+$ branching fractions measured
relative to $\Xi_c^+ \to \Xi^- \pi^+ \pi^+$ to set the absolute scale for many $\Xi_c^+$ branching fractions.

\end{abstract}


\maketitle

\tighten

{\renewcommand{\thefootnote}{\fnsymbol{footnote}}}
\setcounter{footnote}{0}


In recent decades there has been significant experimental progress on the measurements of the weak decays of charmed baryons~\cite{PDG}.
However, given the limited knowledge of
the large nonperturbative effects of quantum chromodynamics, it is difficult to reliably calculate the decay amplitudes of charmed baryons from first principles.
Furthermore, in exclusive charmed-baryon decays the heavy quark expansion does not work. Hence experimental data are needed to extract the nonperturbative quantities in the decay amplitudes~\cite{input1,input2,input3,input4}
and to provide important information to constrain phenomenological models of such decays~\cite{QCD-theory1,QCD-theory2,QCD-theory3,QCD-theory4,QCD-theory5,QCD-theory6,QCD-theory7,QCD-theory8}.

During last few years, Belle and BESIII have measured absolute branching fractions of the $\Lambda_c^+$ and $\Xi_{c}^{0}$ charmed baryons~\cite{lc_belle,lc_bes,myxic}. However, the absolute branching fraction of the remaining member of the charmed-baryon SU(3) flavor antitriplet, the $\Xi_c^+$, has not been measured.  Branching fractions of $\Xi_c^+$ decays have been measured relative to the $\Xi^-\pi^+\pi^+$ mode.
A measurement of the absolute branching fraction $\BR(\Xi_c^+ \to \Xi^- \pi^+ \pi^+)$
is needed to infer the absolute branching fractions of other $\Xi_c^+$ decays.
The comparison of $\Xi_c^+$ decays with those of $\Lambda_c^+$ and $\Xi_c^0$ can also provide an important test of SU(3) flavor symmetry~\cite{Savage:1989qr}.

Along with the reference mode $\Xi_c^+ \to \Xi^- \pi^+ \pi^+$,  $\Xi_c^+ \to p K^- \pi^+$ is a particularly important decay mode as it is the one most often used to reconstruct $\Xi_c^+$ candidates at hadron collider experiments, such as LHCb. For example, the decay has been used to study the properties of $\Xi_b$ and to search for higher excited $\Xi_b$ states via $\Xi_b^0\to \Xi_c^+\pi^-$ \cite{LHCB1,LHCB2}, to search for new $\Omega_c^*$ states in the $\Xi_c^+K^-$ mode \cite{LHCB4}, to measure the doubly charmed baryon via $\Xi_{cc}^{++}\to \Xi_c^+\pi^+$ \cite{LHCB5}, as well as to measure the ratio of fragmentation fractions of $b\to \Xi_b^0$ relative to $b\to \Lambda_b^0$~\cite{LHCB6,LHCB7}.

In experiments, the decay $\Xi_c^+ \to p K^- \pi^+$ has been observed by the FOCUS and SELEX Collaborations and
the branching fraction ratio is measured to be $\BR(\Xi_{c}^{+} \to p K^{-} \pi^{+})/\BR(\Xi_{c}^{+} \to \Xi^{-} \pi^{+} \pi^{+}) = 0.21\pm0.04$~\cite{selex1,focu,selex2,PDG}.
A few models have been developed to predict the decay rates of $\Xi_c^{+}$. For example, the $\BR(\Xi_c^+ \to \Xi^- \pi^+ \pi^+)$ has been predicted to be $(1.47\pm0.84)\%$
based on the SU(3) flavor symmetry~\cite{geng}. Theory predicts $\BR(\Xi_{c}^{+} \to p K^{-} \pi^{+})$ to be $(2.2 \pm 0.8)\%$ based on the measured ratio ${\cal B}(\Xi_c^+ \to p \bar{K}^{*0})/{\cal B}(\Xi_c^+ \to p K^- \pi^+)$ and the $U$-spin symmetry that relates $\Xi_c^+ \to p \bar{K}^{*0}$ and $\Lambda_c^+ \to \Sigma^+ K^{*0}$~\cite{yu1,LHCB7}.
The decay $\bar{B}^{0} \to \bar{\Lambda}_c^- \Xi_c^+$, which proceeds via a $b \to c \bar{c} s$ transition, has been predicted to have a branching fraction of the order $10^{-3}$~\cite{BXL_theroy}, but there has been no experimental measurement.  The world average of the product branching fraction $\BR(\bar{B}^0 \to \bar{\Lambda}_c^- \Xi_c^+) \BR(\Xi_c^+ \to \Xi^- \pi^+ \pi^+)$ is $ (1.8 \pm 1.8) \times 10 ^{-5}$ with large uncertainty~\cite{PDG,belle-old1, babar-old2}.

In this Letter, we perform an analysis of $\bar{B}^{0} \to \bar{\Lambda}_c^- \Xi_c^+$ with ${\bar{\Lambda}}_c^-$ reconstructed via its $\bar{p} K^+ \pi^-$ decay, and $\Xi_c^+$ reconstructed both inclusively and exclusively via the decay modes $\Xi^- \pi^+ \pi^+$ and $p K^- \pi^+$~\cite{charge-conjugate}. We present first a measurement of the absolute branching fraction for $\bar{B}^{0} \to \bar{\Lambda}_c^- \Xi_c^+$
using a missing-mass technique, which is
explained below.
For this analysis we fully reconstruct the
tag-side $B^0$ decay. We subsequently measure the product branching fractions
$\BR(\bar{B}^{0} \to \bar{\Lambda}_c^- \Xi_c^+)\BR(\Xi_c^+ \to \Xi^- \pi^+ \pi^+)$ and $\BR(\bar{B}^{0} \to \bar{\Lambda}_c^- \Xi_c^+)\BR(\Xi_c^+ \to p K^- \pi^+)$
without reconstructing the
recoiling $B^0$ decay in the event as the signal decays are fully reconstructed.
Dividing these product branching fractions by the result for $\BR(\bar{B}^{0} \to \bar{\Lambda}_c^- \Xi_c^+)$ yields the $\BR(\Xi_c^+ \to \Xi^- \pi^+ \pi^+)$ and $\BR(\Xi_c^+ \to p K^- \pi^+)$.

This analysis is based on the full data sample of 711 fb$^{-1}$
collected at the $\Upsilon(4S)$ resonance by the Belle detector~\cite{Belle} at
the KEKB asymmetric-energy $e^{+}e^{-}$ collider~\cite{KEKB}.
To determine detection efficiency and optimize signal event selections, $B$ meson decay events are generated using {\sc evtgen}~\cite{evtgen} and $\Xi_c^+$ inclusive decays are generated using {\sc pythia}~\cite{pythia}. The events are then processed by a detector simulation based on {\sc geant3}~\cite{geant3}. Monte Carlo (MC) simulated samples of $\Upsilon(4S)\to B \bar{B}$ events with $B=B^+$ or $B^0$, and $e^+e^- \to q \bar{q}$ events with $q=u,~d,~s,~c$ at a center-of-mass energy of $\sqrt{s}=10.58$ GeV are used to examine possible peaking backgrounds.



Selection of signal and $\Lambda \to p \pi^-$ candidates uses well reconstructed
tracks and particle identification as described in Ref.~\cite{liyb}.
For the inclusive analysis of the $\Xi_c^+$ decay,
the tag-side $B^0$ meson candidate, $B^0_{\rm tag}$,
is reconstructed using a neural network based on a
full hadron-reconstruction algorithm~\cite{FR}.
Each $B^{0}_{\rm tag}$ candidate has an associated output value ${O}_{\rm NN}$ from the multivariate analysis,
which ranges from 0 to 1. A candidate with larger ${O}_{\rm NN}$ is more likely to be a true $B^0$ meson. If multiple $B_{\rm tag}^0$ candidates are found in an event, the candidate with the largest ${O}_{\rm NN}$ value is selected.
To improve the purity of the $B^{0}_{\rm tag}$ sample, we require ${O}_{\rm NN} > 0.005$, $M_{\rm bc}^{\rm tag} > 5.27$~GeV/$c^{2}$, and $|\Delta E^{\rm tag}|<0.04$~GeV, where
the latter two intervals correspond to approximately 3~standard
deviations, $3\sigma$.
$M_{\rm bc}^{\rm tag}$ and $\Delta E^{\rm tag}$ are defined as
$M_{\rm bc}^{\rm tag} \equiv
\sqrt{E_{\rm beam}^{2} - (\sum_i  \overrightarrow{p}^{\rm
tag}_{\!\! i})^2}$
 and $\Delta E^{\rm tag} \equiv \sum_i E^{\rm tag}_{i}
- E_{\rm beam}$, where $E_{\rm beam}\equiv \sqrt{s}/2$ is the beam energy,
$(E^{\rm tag}_{i}, \overrightarrow{p}^{\rm tag}_{\!\!i})$
is the four-momentum of the $B_{\rm tag}^0$ daughter $i$ in the $\EE$
center-of-mass system (CMS).
${\bar{\Lambda}_c^-} \to \bar{p} K^+ \pi^-$ candidates are selected using the same
method as in Ref.~\cite{myxic}.
A $3 \sigma$ $\bar{\Lambda}_c^-$ signal region is defined by $|M_{\bar{\Lambda}_c^-} - m_{\bar{\Lambda}_c^-}| < 10$~MeV/$c^{2}$. Here and throughout the text, $M_{i}$ represents a measured invariant mass and $m_{i}$ denotes the nominal mass of the particle~$i$~\cite{PDG}.

The mass recoiling against the $\bar{\Lambda}_{c}^{-}$ in $\bar{B}^{0} \to \bar{\Lambda}_c^- + X$
is calculated using $M^{\rm recoil}_{B_{\rm tag}^{0}\bar{\Lambda}_{c}^{-}} = \sqrt{(P_{\rm CMS} - P_{B^{0}_{\rm tag}} - P_{\bar{\Lambda}_{c}^{-}})^{2}}$.
To improve the recoil-mass resolution we use $M_{B_{\rm tag}^{0}\bar{\Lambda}_{c}^{-}}^{\rm rec}\equiv M^{\rm recoil}_{B_{\rm tag}^{0}\bar{\Lambda}_{c}^{-}} + M_{B_{\rm tag}^{0}} - m_{B^{0}} + M_{\bar{\Lambda}_{c}^{-}} -m_{\bar{\Lambda}_{c}^{-}}$. Here, $P_{\rm CMS},~P_{B_{\rm tag}^{0}}$, and $P_{\bar{\Lambda}_{c}^{-}}$ are four-momenta of the initial $\EE$ system, the tagged $B^0$ meson, and the reconstructed $\bar{\Lambda}_{c}^{-}$ baryon, respectively.

\begin{figure}[htbp]
    \begin{center}
        \includegraphics[height=3.25cm]{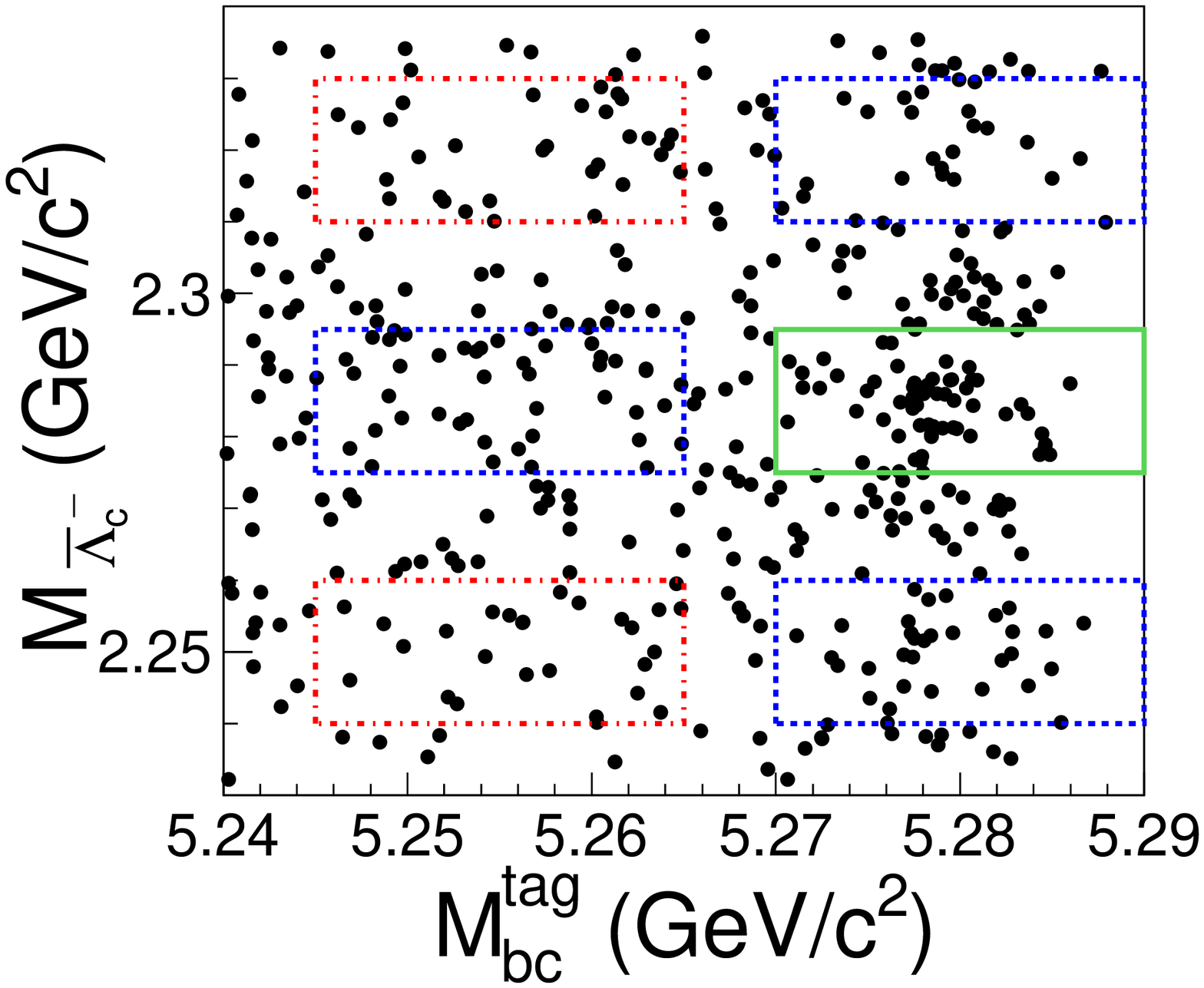}
        \includegraphics[height=3.25cm]{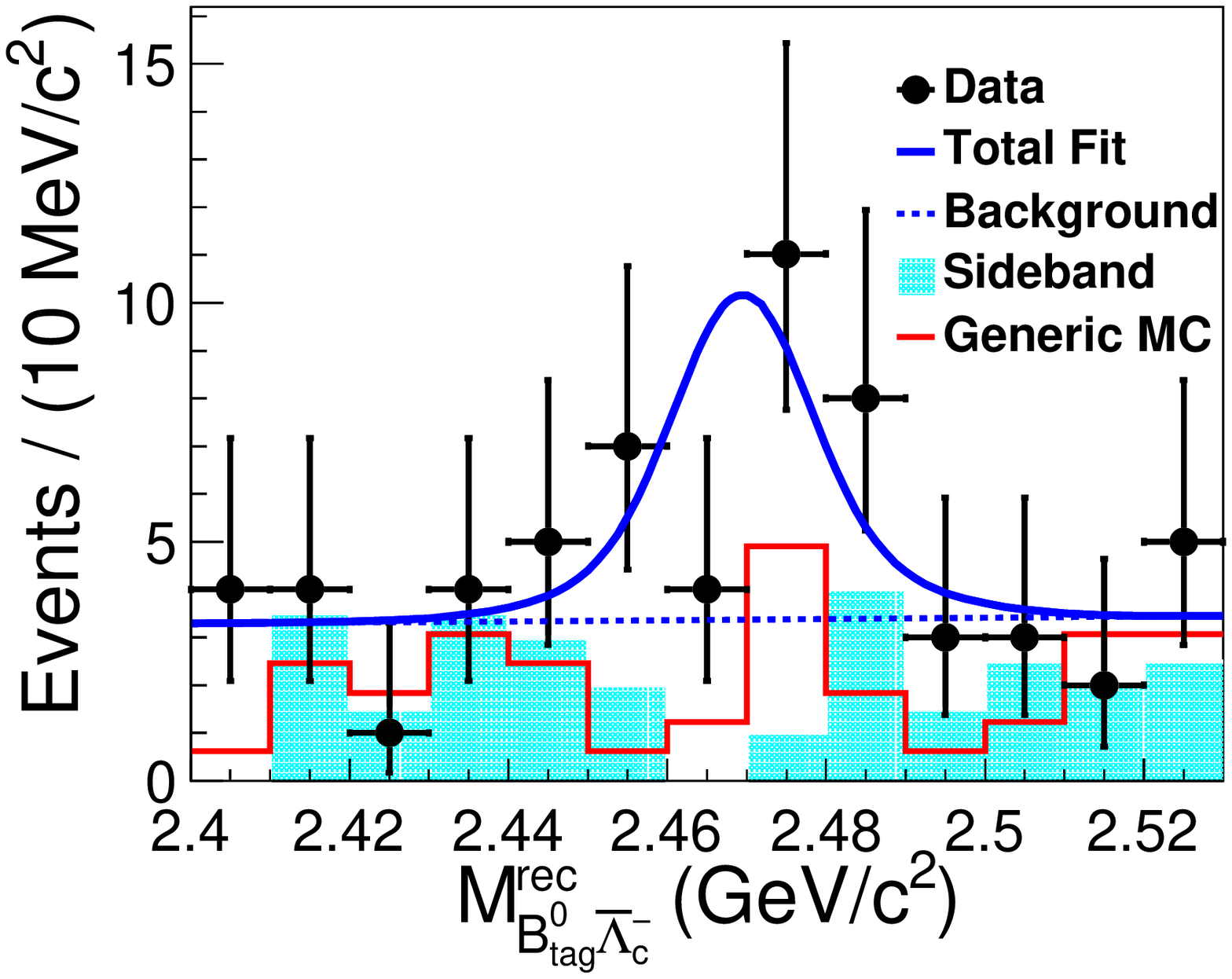}
\caption{The distribution of $M_{\rm bc}^{\rm tag}$ of the $B_{\rm tag}^0$ versus $M_{\bar{\Lambda}_c^-}$ for selected $\bar{B}^{0} \to \bar{\Lambda}_c^- \Xi_c^+$ candidates with $\Xi_c^+ \to anything$ and $\bar{\Lambda}_c^- \to \bar{p} K^{-} \pi^{+}$ (left) and the fit to the $M_{B_{\rm tag}^0\bar{\Lambda}_c^-}^{\rm rec}$ distribution (right).
The solid box shows the selected signal region. The blue dashed and red dash-dotted boxes define the $M_{\rm bc}^{\rm tag}$ and $M_{\bar{\Lambda}_c^-}$ sidebands described in the text.
The points with error bars are the data in the signal box, the solid blue curve is the best fit, the dashed curve is the fitted background, the cyan shaded histogram is the normalized $M_{\rm bc}^{\rm tag}$ and $M_{\bar{\Lambda}_c^-}$ sidebands, the red open histogram is the sum of the MC-simulated contributions for $\EE \to q\bar{q}$, and $\Upsilon(4S) \to B \bar{B}$ generic-decay backgrounds with the number of events normalized to the number of events from the normalized $M^{\rm tag}_{\rm bc}$ and $M_{\bar{\Lambda}_c^-}$ sidebands.
}\label{FR_mbc_mlc_com}
    \end{center}
\end{figure}

Figure ~\ref{FR_mbc_mlc_com} (left) shows the distribution of $M_{\rm bc}^{\rm tag}$ of the $B^{0}_{\rm tag}$ candidates versus $M_{\bar{\Lambda}_c^-}$ of the selected $\bar{B}^{0} \to \bar{\Lambda}_c^- \Xi_c^{+}$ signal candidates after all selection requirements in the studied $\Xi_c^+$ mass region of $2.4<M_{B_{\rm tag}^{0}\bar{\Lambda}_{c}^{-}}^{\rm rec}<2.53$ GeV/$c^2$.
Candidates $\bar{B}^{0} \to \bar{\Lambda}_c^- \Xi_c^+$ are observed in the signal region defined by the solid box. To check possible peaking backgrounds, we define $M_{\rm bc}^{\rm tag}$ and $M_{\bar{\Lambda}_c^-}$ sidebands, which are represented by the dashed and dash-dotted boxes.
The normalized contribution of the $M_{\rm bc}^{\rm tag}$ and $M_{\bar{\Lambda}_c^-}$ sidebands is estimated
as being half the number of events in the blue dashed boxes minus one fourth the number of events in
the red dash-dotted boxes. The $M_{B_{\rm tag}^{0}\bar{\Lambda}_{c}^{-}}^{\rm rec}$ distribution in the signal and the sideband boxes is shown in Figure~\ref{FR_mbc_mlc_com} (right).

To extract the $\Xi_c^+$ signal yields we perform an unbinned maximum likelihood fit to the $M_{B_{\rm tag}^0\bar{\Lambda}_c^-}^{\rm rec}$ distribution. A double-Gaussian function with its parameters fixed to those from a fit to the MC-simulated signal distribution
is used to model the $\Xi_c^+$ signal shape and a first-order polynomial is used for the background shape
since we find no peaking background in the $M_{\rm bc}^{\rm tag}$ and $M_{\bar{\Lambda}_c^-}$ sideband events.
For all the fits described in this paper, the signal and background yields, and the parameters of the
background shape are left free.
The fit results are shown in Figure~\ref{FR_mbc_mlc_com} (right).

The fitted number of $\Xi_c^+$ signal events is $N_{\Xi_c^+} = 18.8 \pm 6.8$. This corresponds to a statistical significance of $3.2\sigma$ estimated using $\sqrt{-2\ln(\mathcal{L}_{0}/\mathcal{L}_{\rm max})}$, where $\mathcal{L}_{0}$ and $\mathcal{L}_{\rm max}$ are the maximum likelihood values of the fits without and with a signal component, respectively.
The signal significance becomes 3.1$\sigma$ once we convolve
the likelihood with a Gaussian function whose width equals the total systematic uncertainty.
The signal significance found using alternative fits to the $M_{B_{\rm tag}^0\bar{\Lambda}_c^-}^{\rm rec}$ distribution as described in the section on systematic uncertainties, is greater than 3.0$\sigma$ in all cases.
The branching fraction is \begin{align*} \BR(\bar{B}^{0} \to \bar{\Lambda}_c^- \Xi_c^+) &= N_{\Xi_c^+}/ [2 N_{\bar{B}^0} \varepsilon_{\rm inc} \BR(\bar{\Lambda}_c^- \to \bar{p} K^+ \pi^-)]\\  &= [1.16 \pm 0.42(\rm stat.)] \times 10^{-3},\end{align*} where $N_{\bar{B}^0}=N_{\Upsilon(4S)}\BR(\Upsilon(4S)\to B^0\bar{B}^0)$, $N_{\Upsilon(4S)}$ is the number of $\Upsilon(4S)$ events, and $\BR(\Upsilon(4S)\to B^0\bar{B}^{0})=0.486$~\cite{PDG}. The reconstruction efficiency, $\varepsilon_{\rm inc}$, is obtained from the MC simulation.
The $\BR(\bar{\Lambda}_c^- \to \bar{p} K^+ \pi^-)$ is taken from Ref.~\cite{PDG}.


For the analysis of the exclusive $\Xi_c^+$ decays, we
reconstruct $\Xi_c^+$ from $\Xi^- \pi^+ \pi^+$ with $\Xi^- \to \Lambda \pi^-\ (\Lambda \to p \pi^-)$ and $\Xi^- \to p K^- \pi^+ $ modes, with no $B^0_{\rm tag}$. The daughters of the $\bar{B}^{0}$, $\Xi_c^+$, and $\Xi^-$ candidates are
fit to common vertices. If there is more than one $\bar{B}^{0}$ candidate in an event, the one with the smallest $\chi^{2}_{\rm vertex}/{\rm n.d.f.}$ from the $\bar{B}^{0}$ vertex fit is selected.
The requirements of $\chi^{2}_{\rm vertex}/{\rm n.d.f.} <50,~15$, and $15$ are applied to reconstructed $\bar{B}^{0}$, $\Xi_c^+$, and $\Xi^-$ candidates, respectively, with selection efficiencies above 96\%, 95\%, and 95\%.  $\Xi^-$ and $\Xi_c^+$ signals are defined as $|M_{\Xi^-} - m_{\Xi^-}| < 10$~MeV/$c^2$ and $|M_{\Xi_c^+} - m_{\Xi_c^+}| < 20$~MeV/$c^2$ corresponding to about $3 \sigma$.
The $\bar{\Lambda}_c^-$ signal interval is the same as in the inclusive analysis of $\Xi_c^+$ decays.
$\bar{B}^{0}$ signal candidates are identified using the beam-constrained mass $M_{\rm bc}$ and the energy difference $\Delta E$. Here, $M_{\rm bc}$ and $\Delta E$ are defined as $M_{\rm bc}^{\rm tag}$ and $\Delta E^{\rm tag}$ above,
but calculated using the momenta of the signal candidate tracks directly.

\begin{figure*}[htbp]
    \begin{center}
        \includegraphics[width=5cm,height=3.65cm]{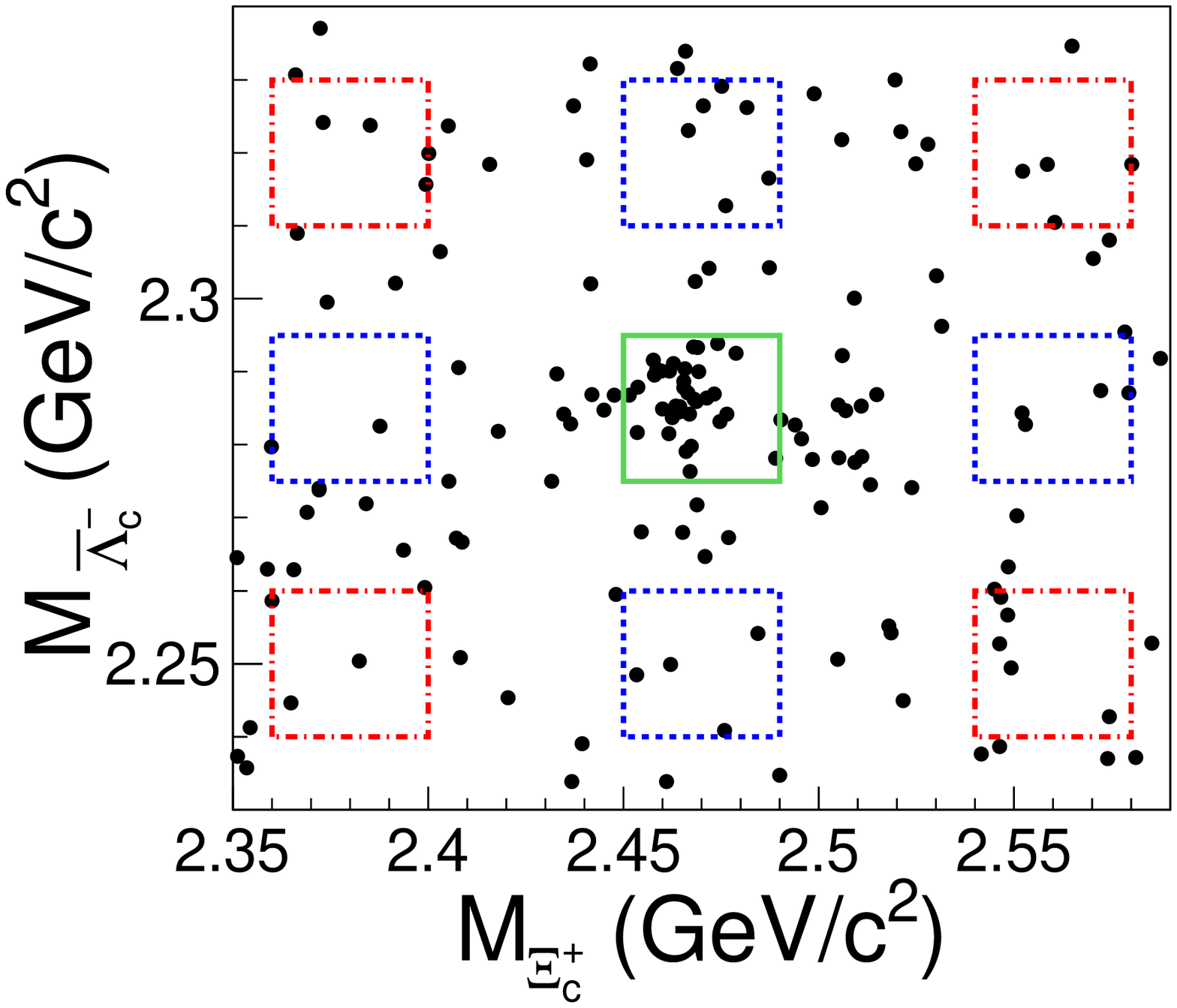}
        \includegraphics[width=5cm]{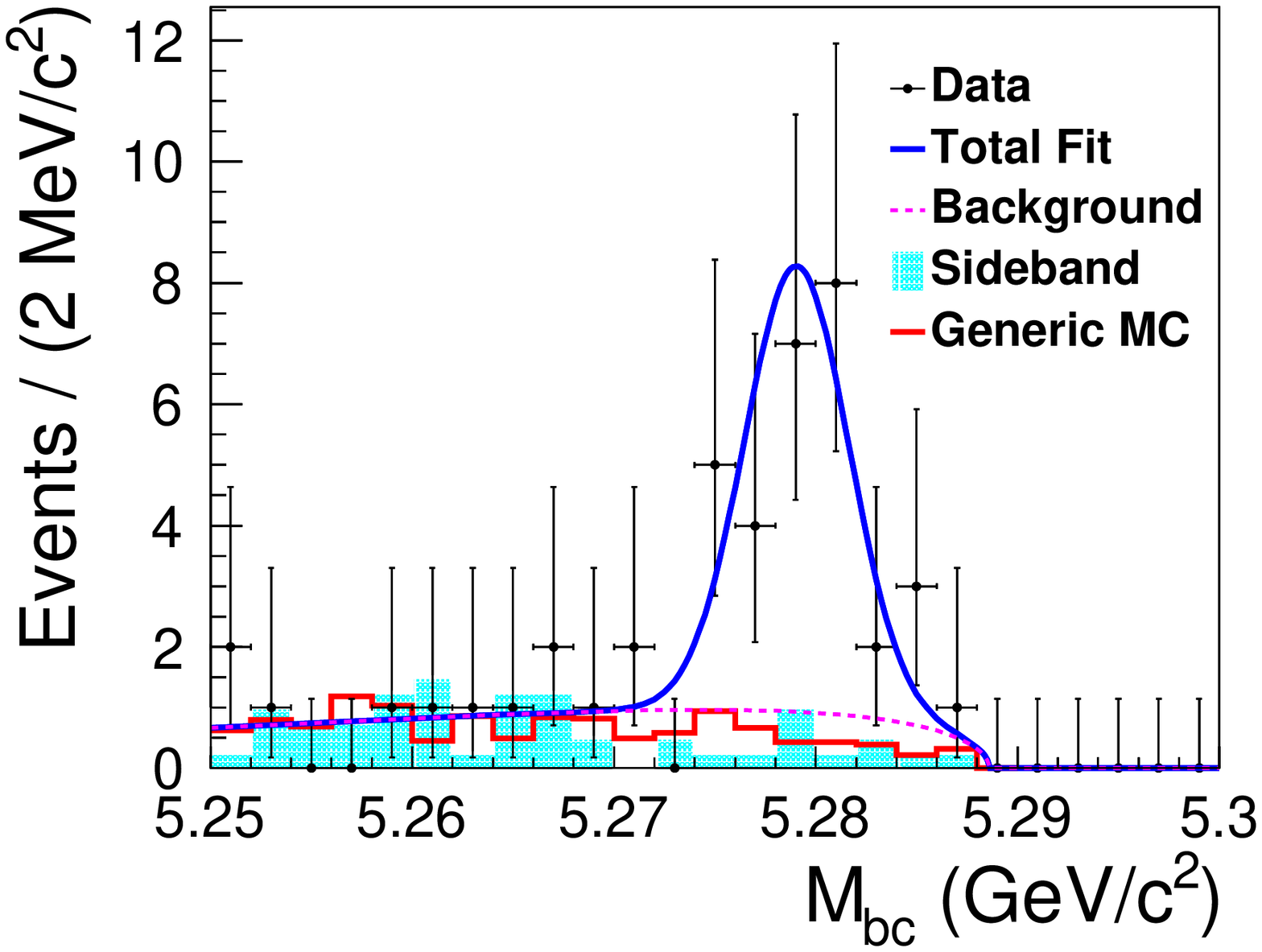}
        \includegraphics[width=5cm]{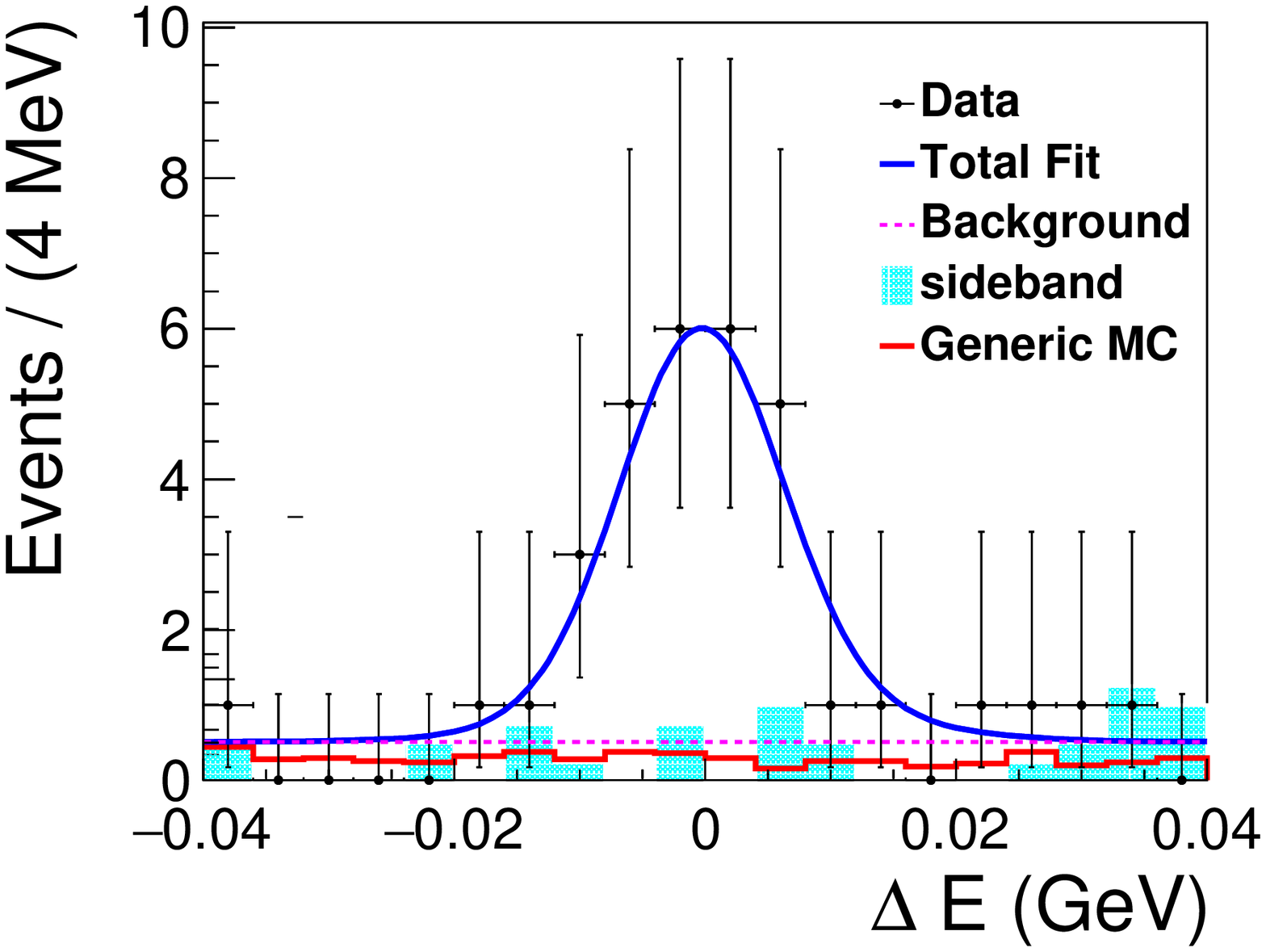}
     \textcolor{red}{\put(-385,79){\bf (a1)} \put(-250,79){\bf (b1)}  \put(-100,79){\bf (c1)}}

        \includegraphics[width=5cm,height=3.65cm]{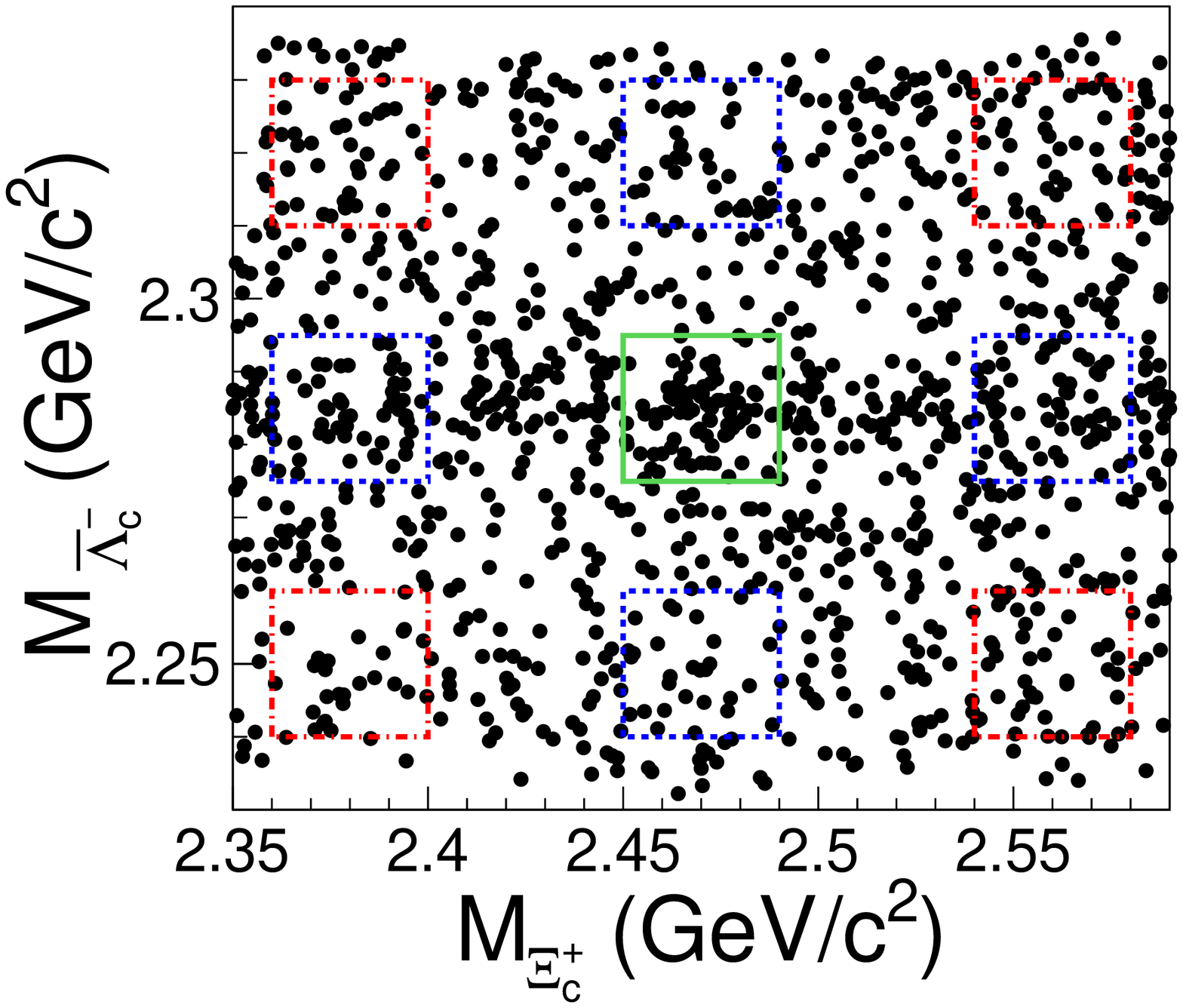}
        \includegraphics[width=5cm]{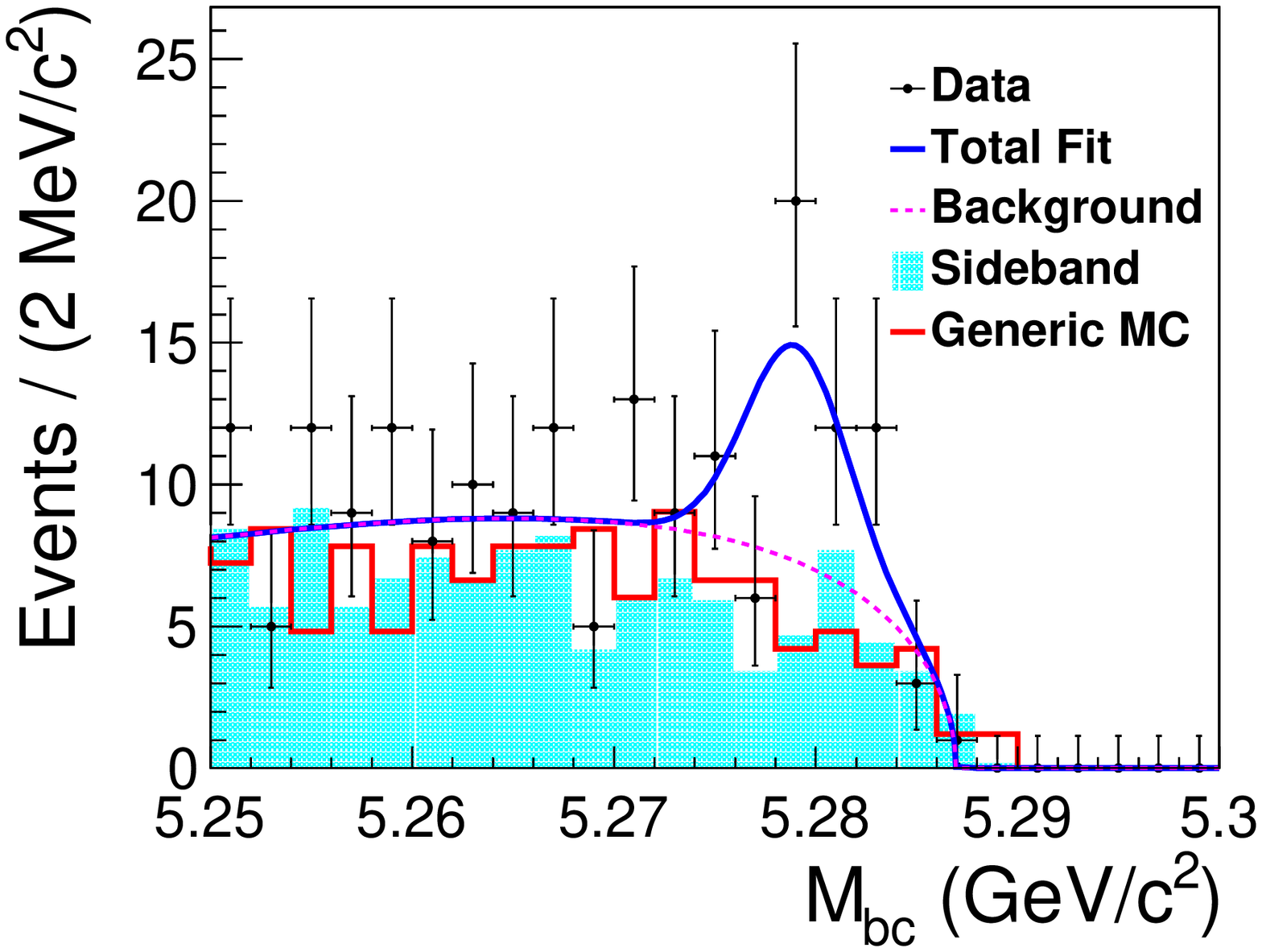}
        \includegraphics[width=5cm]{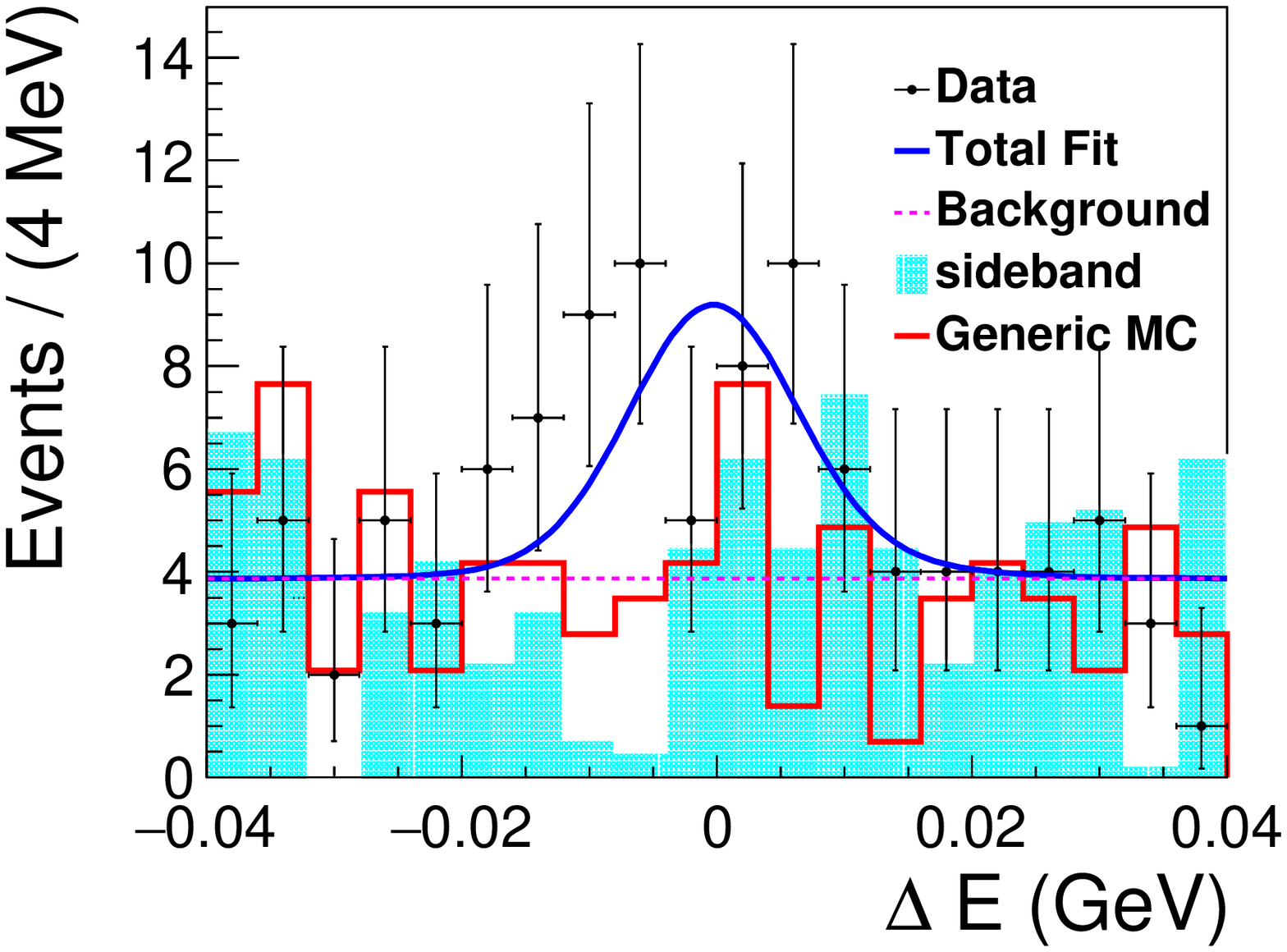}
     \textcolor{red}{\put(-385,79){\bf (a2)} \put(-250,79){\bf (b2)}  \put(-100,79){\bf (c2)}}


\caption{The distributions of (a) $M_{\Xi_c^+}$ versus $M_{\bar{\Lambda}_c^-}$, and the fits to the (b) $M_{\rm bc}$ and (c) $\Delta E$ distributions of the selected $\bar{B}^{0} \to \bar{\Lambda}_c^- \Xi_c^+$ candidates for (b1-c1) the $\Xi_c^+ \to \Xi^- \pi^{+} \pi^{+}$ and (b2-c2) the $\Xi_c^+ \to p K^- \pi^+$ decay modes.
In plots (a1-a2), the central solid boxes are the signal regions, and the red dash-dotted and blue dashed boxes show the $M_{\Xi_c^+}$ and $M_{\bar{\Lambda}_c^-}$ sidebands used to estimate of the backgrounds (see text). The dots with error bars are the data, the blue solid curves represent the best fits, and the dashed curves represent the fit background contributions.  The shaded histograms are the normalized as in the text $M_{\Xi_c^+}$ and $M_{\bar{\Lambda}_c^-}$ sidebands, the red open histograms
represent the generic background described in
the caption of Fig.~\ref{FR_mbc_mlc_com}.} \label{mxic_mlc_data_com}
    \end{center}
\end{figure*}

After the event selections, the distributions of $M_{\Xi_c^+}$ versus $M_{\bar{\Lambda}_c^-}$ in the $\bar{B}^{0}$ signal region defined by $|\Delta E| < 0.03\,~{\rm GeV}$ and $M_{\rm bc} > 5.27\,{\rm GeV}/c^2$ corresponding to about $3 \sigma$ are shown in Figures~\ref{mxic_mlc_data_com}(a1) and \ref{mxic_mlc_data_com}(a2). The central solid boxes are the $\Xi_c^+$ and $\bar{\Lambda}_c^-$ signal regions. The backgrounds from non-$\Xi_c^+$ and non-$\bar{\Lambda}_c^-$ events are estimated with the $M_{\Xi_c^+}$ and $M_{\bar{\Lambda}_c^-}$ sidebands, represented by the dashed and dash-dotted boxes in Figures~\ref{mxic_mlc_data_com}(a1) and \ref{mxic_mlc_data_com}(a2).
The normalized contributions from the $M_{\Xi_c^+}$ and $M_{\bar{\Lambda}_c^-}$ sidebands are estimated
using half the number of events in the blue dashed boxes minus one fourth the number of events in the red dash-dotted boxes.
Figure~\ref{mxic_mlc_data_com} shows the $M_{\rm bc}$ and $\Delta E$ distributions in the $\Xi_c^+$ and $\bar{\Lambda}_c^-$ signal regions from the selected $\bar{B}^{0} \to \bar{\Lambda}_c^- \Xi_c^+$ candidates with  $\Xi_c^+ \to \Xi^- \pi^+ \pi^{+}$ (b1-c1) and  $\Xi_c^+ \to p K^- \pi^+$ (b2-c2) decay modes.

We perform a two-dimensional (2D) maximum likelihood fit to the $M_{\rm bc}$ and $\Delta E$ distributions to extract the number of
$\bar{B}^{0} \to \bar{\Lambda}_c^- \Xi_c^+$ signal events with $\Xi_c^+ \to \Xi^- \pi^{+} \pi^{+}/p K^- \pi^+$. For the $M_{\rm bc}$ distribution, the signal shape is modeled using a Gaussian function and the background is described using an ARGUS function~\cite{argus}.  For the $\Delta E$ distribution, the signal shape is a double-Gaussian and the background is a first-order polynomial. All shape parameters of the signal functions are fixed to the values obtained from the fits to the MC simulated signal distributions. The fit results are shown in Figure~\ref{mxic_mlc_data_com}.

The signal yields are $N_{\Xi^- \pi^{+} \pi^{+}}= 24.2 \pm  5.4$ (6.9$\sigma$ significance
and 6.8$\sigma$ with systematic uncertainties included) and $N_{p K^- \pi^+} = 24.0\pm  6.9$ (4.5$\sigma$ significance and 4.4$\sigma$ with systematic uncertainties included).  We use the efficiencies from MC simulations to
measure $\BR(\bar{B}^{0} \to \bar{\Lambda}_c^- \Xi_c^+) \BR(\Xi_c^+ \to \Xi^- \pi^{+} \pi^+)$
and $\BR(\bar{B}^{0} \to \bar{\Lambda}_c^- \Xi_c^+) \BR(\Xi_c^+ \to p K^- \pi^+)$ as $[3.32 \pm 0.74(\rm stat.)] \times 10^{-5}$
and $[5.27 \pm 1.51(\rm stat.)] \times 10^{-6}$, respectively.

We divide the above product branching fractions by the value of $\BR(\bar{B}^{0} \to \bar{\Lambda}_c^- \Xi_c^+)$ and
for the first time measure $\BR(\Xi_c^+ \to \Xi^- \pi^{+} \pi^+)$,
$\BR(\Xi_c^+ \to p K^- \pi^+)$, and the ratio
between them.  These are listed in Table~\ref{tab:err}.


There are several sources of systematic uncertainties in the branching fraction measurements.
The uncertainties related to reconstruction efficiency include those for tracking efficiency (0.35\% per track),
particle identification efficiency (0.9\% per kaon, 0.9\% per pion, and
3.3\% per proton), as well as $\Lambda$  reconstruction efficiency (3.0\% per $\Lambda$~\cite{lambda}).
We assume these reconstruction-efficiency-related uncertainties are independent and sum them in
quadrature.
We estimate the systematic uncertainties associated with the fitting procedures by changing the order of the background polynomial, the range of the fit, and by enlarging the mass resolution by 10\%. The observed deviations from the nominal fit results are taken as systematic uncertainties. The uncertainty on $\BR(\bar{\Lambda}_c^- \to \bar{p} K^+ \pi^-)$ is taken from Ref.~\cite{PDG}.
The uncertainty due to the $B^{0}$ tagging efficiency is  4.5\%~\cite{FRerr}.
A relative systematic uncertainty on $\BR(\Upsilon(4S)\to B^{0}\bar{B}^0)$ is 1.23\%~\cite{PDG}.
The systematic uncertainty on $N_{\Upsilon(4S)}$ is 1.37\%~\cite{n4serr}.  For the $\Xi_c^+$ branching fractions and the corresponding ratio, some common systematic uncertainties, including tracking, particle identification, $\bar{\Lambda}_c^-$ decay branching fraction, $\Lambda$ selection, and the total number of $B\bar{B}$ pairs, cancel.
We summarize the sources of systematic uncertainties in Table~\ref{tab:err}, assume them to be independent, and
add them in quadrature to obtain the total systematic uncertainties.

\begin{table*}[htbp]
\caption{\label{tab:err} Summary of the measured $\Xi_c^+$ branching fractions and ratio (last column), and the corresponding systematic uncertainties in \%. For the branching fractions and ratio, the first uncertainties are statistical and the second are systematic.}
    \begin{tabular}{l||c|c|c|c|c|c||c}
        \hline\hline
        Observable & Efficiency  & Fit  &  $\Lambda_c$ decays & $B_{\rm tag}$ &\tabincell{c}{$N_{\bar{B}^{0}}$}  & \tabincell{c}{Sum} &  Measured value\\
        \hline
        $\BR(\bar{B}^{0} \to \bar{\Lambda}_c^- \Xi_c^+)$ & 3.66 & 10.3  & 5.3 & 4.5 & 1.82 & 13.1 & $(1.16 \pm 0.42 \pm 0.15)\times 10^{-3}$ \\
        $\BR(\bar{B}^{0} \to \bar{\Lambda}_c^- \Xi_c^+)\BR(\Xi_c^+ \to \Xi^- \pi^+\pi^+)$ &6.24 &5.61 &5.3  &$...$  & 1.82& 10.1 & $(3.32 \pm 0.74 \pm 0.33) \times 10^{-5}$ \\
        $\BR(\bar{B}^{0} \to \bar{\Lambda}_c^- \Xi_c^+) \BR(\Xi_c^+ \to p K^- \pi^+)$  &7.32  & 9.53 &5.3 &$...$   &1.82  &13.3 & $(5.27 \pm 1.51 \pm 0.69)\times 10^{-6}$ \\
        \hline
        $\BR(\Xi_c^+ \to \Xi^- \pi^+ \pi^+)$ & 4.23 & 11.7  &$...$ & 4.5 & $...$& 13.2 & $(2.86 \pm 1.21 \pm 0.38)\%$ \\
        $\BR(\Xi_c^+ \to p K^- \pi^+)$ & 3.66 & 14.0 & $...$ & 4.5 & $...$  & 15.2 & $(0.45 \pm 0.21 \pm 0.07)\%$ \\\hline
        $\BR(\Xi_c^+ \to p K^- \pi^+) /\BR(\Xi_c^+ \to \Xi^- \pi^{+} \pi^+)$ &4.90 & 11.0 & $...$ & $...$ & $...$  & 12.0  & $0.16 \pm 0.06 \pm 0.02$ \\
        \hline\hline
    \end{tabular}
\end{table*}

We report the first measurements of the absolute branching fractions
\begin{displaymath}
\begin{aligned}
\BR(\Xi_c^+ \to \Xi^- \pi^{+} \pi^+)=(2.86 \pm 1.21\pm 0.38)\%, \\
\BR(\Xi_c^+ \to p K^- \pi^+)=(0.45 \pm 0.21\pm 0.07)\%,\\
\end{aligned}
\end{displaymath}
where the first uncertainties are statistical and the second systematic.
The measured $\BR(\Xi_c^+ \to \Xi^- \pi^{+} \pi^+)$ value is consistent with
the theoretical prediction within uncertainties~\cite{geng}.
The measured central value of $\BR(\Xi_c^+ \to p K^- \pi^+)$
is smaller than that of the theoretical predictions~\cite{yu1,LHCB7}, perhaps
indicating a large $U$-spin symmetry breaking effect in the singly-Cabibbo-suppressed
charmed-baryon decays.
The branching fraction $\BR(\bar{B}^{0} \to \bar{\Lambda}_c^- \Xi_c^+)$ is measured for the first time to be $[1.16 \pm 0.42(\rm stat.) \pm 0.15(\rm syst.)] \times 10^{-3}$ and agrees well with that of $B^- \to \bar{\Lambda}_c^- \Xi_c^0$~\cite{myxic} which is consistent with the  expectation from isospin symmetry.
The product branching fractions are \begin{align*}\BR(\bar{B}^{0} \to \bar{\Lambda}_c^- \Xi_c^+)\BR(\Xi_c^+ \to \Xi^- \pi^{+} \pi^+)\\= (3.32 \pm 0.74 \pm 0.33) \times 10^{-5},\\ \BR(\bar{B}^{0} \to \bar{\Lambda}_c^- \Xi_c^+) \BR(\Xi_c^+ \to p K^- \pi^+)\\=(5.27 \pm 1.51 \pm 0.69) \times 10^{-6}.\end{align*}
The first of these branching fraction measurements is consistent with previous measurements, with improved precision, and supersedes the Belle measurement~\cite{belle-old1}.
The ratio $\BR(\Xi_c^+ \to p K^- \pi^+)/\BR(\Xi_c^+ \to \Xi^- \pi^{+} \pi^+)$
is measured to be $0.16 \pm 0.06(\rm stat.) \pm 0.02(\rm syst.)$, which is consistent with world-average value of
$0.21~\pm~0.04$~\cite{PDG} within uncertainties.
Our measured $\Xi_c^+$ branching fractions, e.g. for $\Xi_c^+ \to \Xi^-
\pi^+ \pi^+$, can be combined with $\Xi_c^+$ branching fractions measured relative to
$\Xi_c^+ \to \Xi^- \pi^+ \pi^+$ to yield other absolute $\Xi_c^+$ branching fractions.

In summary, based on $(772 \pm 11) \times 10^{6}$ $B\bar{B}$ pairs collected at the $\Upsilon(4S)$ resonance with the Belle detector,
we perform an analysis of $\bar{B}^{0} \to \bar{\Lambda}_c^- \Xi_c^+$
inclusively
using a hadronic $B$-tagging method based on a full reconstruction algorithm~\cite{FR},
and exclusively with $\Xi_c^+$ decays into $\Xi^- \pi^{+} \pi^+$ and $p K^- \pi^+$ final states.
These are the first measurements of the absolute branching fractions $\BR(\Xi_c^+ \to \Xi^- \pi^{+} \pi^+)$  and $\BR(\Xi_c^+ \to p K^- \pi^+)$.

We thank Professor Fu-sheng Yu for useful discussions and comments.
We thank the KEKB group for excellent operation of the
accelerator; the KEK cryogenics group for efficient solenoid
operations; and the KEK computer group, the NII, and
PNNL/EMSL for valuable computing and SINET5 network support.
We acknowledge support from MEXT, JSPS and Nagoya's TLPRC (Japan);
ARC (Australia); FWF (Austria); NSFC and CCEPP (China);
MSMT (Czechia); CZF, DFG, EXC153, and VS (Germany);
DST (India); INFN (Italy);
MOE, MSIP, NRF, RSRI, FLRFAS project and GSDC of KISTI and KREONET/GLORIAD (Korea);
MNiSW and NCN (Poland); MSHE (Russia); ARRS (Slovenia);
IKERBASQUE (Spain);
SNSF (Switzerland); MOE and MOST (Taiwan); and DOE and NSF (USA).


\begin{thebibliography}{99}
\bibitem{charge-conjugate} Inclusion of charge-conjugate states is implicit unless otherwise stated.
\end{thebibliography}
\end{document}